# An Effective Machine-Part Grouping Algorithm to Construct Manufacturing Cells

Tamal Ghosh[*1], Pranab K Dan[2]

*Department of Industrial Engineering & Management, West Bengal University of Technology Kolkata.*
*BF 142, Salt Lake City, Kolkata 700064, India.*
*Tel. +91-33-2334-1014; Fax: +91-33-2334-1030*
[1]`tamal.31@gmail.com`
[2]`danpk.wbut@gmail.com`
[*]*Corresponding Author*

*Abstract*—The machine-part cell formation problem consists of creating machine cells and their corresponding part families with the objective of minimizing the inter-cell and intra-cell movement while maximizing the machine utilization. This article demonstrates a hybrid clustering approach for the cell formation problem in cellular manufacturing that conjoins Sorenson's similarity coefficient based method to form the production cells. Computational results are shown over the test datasets obtained from the past literature. The hybrid technique is shown to outperform the other methods proposed in literature and including powerful soft computing approaches such as genetic algorithms, genetic programming by exceeding the solution quality on the test problems.

## I. INTRODUCTION

In cellular manufacturing systems (CMS), group technology (GT) could be stated as a manufacturing philosophy which recognises similar parts and clusters them into part families depending upon its manufacturing designs, features and geometric shapes [1]. Designing manufacturing cell is known as cell formation problem (CF/CFP). It consists of the following procedures: usually similar parts are grouped into part families following their processing requirements, and heterogeneous machines are grouped into manufacturing cells and subsequently part families are designated to cells. The problem encountered in CMS is construction of such cells irrespective of its type [2]. Not essentially the aforementioned steps are carried out in the above order or even sequentially. Depending upon the procedures involved in CFP three methods of achieving solutions are proposed [2]: (1) recognizing part families first and consequently machines are clustered into cells depending on the processing requirement of part families, (2) recognizing manufacturing cells by grouping heterogeneous machines and then the part families are allocated to cells, (3) part families and machine cells are developed concurrently. Due to the NP-Complete nature of the problem, many computational techniques are heavily practised for improved solution to the CFP, a thorough discussion can be found in literature [3].

Over the past few decades many hierarchical and non-hierarchical techniques are adopted by researchers in the aforesaid domain such as ZODIAC [4], GRAFICS [5], MST [6], K-Harmonic Mean algorithm [7] etc.

Various other techniques are developed to solve manufacturing cell formation problems since last forty years, these include similarity coefficient methods, clustering analysis, array based techniques, graph partitioning methods etc. The similarity coefficient approach was first suggested by McAuley [8]. The basis of similarity coefficient methods is to calculate the similarity between each pair of machines and then to group the machines into cells based on their similarity measurements. Few studies have proposed to measure dissimilarity coefficients instead of similarity coefficient for machine-part grouping problems [9].

Machine–part grouping problem is based on production flow analysis, in which the machine-part production cells are formed by permuting rows and columns of the machine-part mapping chart in the form of a {0-1} incidence matrix. Some of the methods are Rank order clustering [10], Bond energy algorithm [11] etc. Dimopoulos and Mort proposed a hierarchical algorithm combined with genetic programming for cell formation problem [12].

Array based methods consider the rows and columns of the machine-part incidence matrix as binary patterns and reconfigure them to obtain a block diagonal cluster formation. The rank order clustering algorithm is the most familiar array-based technique for cell formation [10]. Substantial alterations and enhancements over rank order clustering algorithm have also been studied [13, 14]. The direct clustering analysis (DCA) has been stated by Chan and Milner [14], and bond energy analysis is performed by McCormick et al. [11].

Graph Theoretic Approach depicts the machines as vertices and the similarities between machines as the weights on the arcs. Rajagopalan and Batra proposed the use of graph theory to form machine cells [15]. An ideal seed non-hierarchical clustering algorithm for cellular manufacturing stated in [13]. A non-heuristic network method was also stated to construct manufacturing cells with minimum inter-cell moves [16]. Srinivasan implemented a method using minimum spanning tree (MST) for the machine-part cell formation problem [6]. A polynomial-time algorithm based on a graph theoretic approach was developed by Veeramani and Mani, named as vertex-tree graphic matrices [17].

This article presents a hybrid approach based on centroid linkage hierarchical clustering technique which is combined with Sorenson's similarity coefficient method [18] to form the manufacturing cells.

## II. DEFINITION OF PROBLEM

The cell formation problem in group technology begins with two fundamental tasks, namely, machine-cell formation and part-family identification. To form machine-cell, similar machines are grouped to process one or more part-families. In part-family formation, parts with similar design features, attributes, shapes are grouped, so that the group of parts can be manufactured within a machine cell. Generally, the cell formation problems are represented in a matrix namely 'machine-part incident matrix'. Its elements are either 0 or 1. Parts are arranged in columns and machines are in rows in the incidence matrix. An example matrix is presented in Fig. 1.

|    | P1 | P2 | P3 | P4 | P5 |
|----|----|----|----|----|----|
| M1 |    | 1  | 1  |    | 1  |
| M2 | 1  |    |    | 1  | 1  |
| M3 |    |    | 1  |    | 1  |

Fig 1. Machine-part incidence matrix (3×5)

It depicts that machine 1 processes part 2, 3, 5, machine 2 processes part 1, 4 and machine 3 processes part 3, 5. In this matrix a 0 indicates no mapping or no processing and an 1 indicates mapping or processing. The final block diagonal structure is shown in Fig. 2.

|    | P3 | P5 | P2 | P1 | P4 |
|----|----|----|----|----|----|
| M1 | 1  | 1  | 1  | 0  | 0  |
| M3 | 1  | 1  | 0  | 0  | 0  |
| M2 | 0  | 1  | 0  | 1  | 1  |

Fig. 2 Block diagonal matrix (3×5)

In Fig. 2 cells are shown block diagonally as square boxes. cell 1 contains machine 1 and 3 and part 2, 3, 5 and cell 2 contains machine 2 and part 2 and 4. A 1 outside the block means a part is processed through some machine which does not belong to the corresponding machine cell, therefore the intercellular move cost will be added. This element is known as an exceptional element (EE) and a 0 inside a cell means an unutilized space in cell, therefore lesser utilization of space. It is known as 'void'. The objective of cell formation is to minimize the EEs and voids.

## III. PERFORMANCE METRIC

Two widely accepted performance measures to assess the goodness of CFP solutions are grouping efficiency [13] which incorporates machine utilization and intercell moves, and Grouping efficacy [19], intends to minimize the number of exceptional elements and the number of voids in the diagonal blocks. A detailed description regarding various performance measure could be obtained from a critical survey of reference [20]. In this study grouping efficacy measure is used as the solution evaluation. Grouping efficacy measure is stated as,

$$\tau = \frac{E - E_e}{E + E_v} = 1 - \frac{E_v + E_e}{E + E_v} \quad (1)$$

Where
$E$ = Total number of 1s in incidence matrix
$E_e$ = Total number of exceptional elements
$E_v$ = Total number of voids

## IV. METHODOLOGY

Similarity coefficient based techniques are massively practiced in formation of manufacturing cells as found in past literature [21]. In this article the similarity measure method is utilized namely Sorenson's similarity coefficient [18].

$$S_{ij} = \frac{2a_{ij}}{2a_{ij} + b_{ij} + c_{ij}} \quad (2)$$

$S_{ij}$ = Similarity between machine $i$ and machine $j$,
$a_{ij}$ = the number of parts processed by both machines $i$ and $j$,
$b_{ij}$ = the number of parts processed by machine $i$ but not by machine $j$,
$c_{ij}$ = the number of parts processed by machine $j$ but not by machine $i$.

Centroid Linkage Clustering Algorithm is adopted in this study as the solution methodology which is theoretically and mathematically simple algorithm practiced in hierarchical agglomerative clustering analysis of mixed data [22]. It delivers informative descriptions and visualization of possible data clustering structures. When there exists hierarchical relationship in data this approach can be more competent.

### A. input dataset

An input data set is a machine–part incidence matrix. Machines are the items that should be grouped based on their similarities. Parts are the components which contains routing information. An example input dataset of past literature [23] is presented in Fig. 3

|    | p1 | p2 | p3 | p4 | p5 | p6 | p7 |
|----|----|----|----|----|----|----|----|
| m1 |    | 1  |    | 1  | 1  | 1  |    |
| m2 | 1  |    | 1  |    |    |    |    |
| m3 | 1  |    | 1  |    |    | 1  | 1  |
| m4 |    | 1  |    | 1  |    | 1  |    |
| m5 | 1  |    |    |    | 1  |    | 1  |

Fig. 3 Incidence input matrix (5×7) [23]

### B. Calculating similarity value between machines

This function computes the similarity between each pairs of machines of the given input of data matrix using the equation (2). It produces a similarity matrix separately for machines. The similarity matrix of Fig. 3 is presented in Fig. 4.

|    | m1   | m2   | m3   | m4 | m5 |
|----|------|------|------|----|----|
| m1 | 1    |      |      |    |    |
| m2 | 0    | 1    |      |    |    |
| m3 | 0.25 | 0.67 | 1    |    |    |
| m4 | 0.86 | 0    | 0.29 | 1  |    |
| m5 | 0.29 | 0.4  | 0.57 | 0  | 1  |

Fig.4 Similarity matrix of machines

## C. Dendogram Formation

The proposed technique takes the input as a similarity matrix from the previous step and produces dendrogram structure that links individual machines or subgroup of machines according to their values of similarity coefficients. Centroid linkage function is implemented on the basis of hierarchical cluster information. If cell $r$ is formed from cell $p$ and $q$, and $n_r$ is the number of machines in cell $r$, $x_{ri}$ is the $i^{th}$ machine of cell $r$, then centroid linkage is computed using the formula,

$$d(r,s) = \|\bar{x}_r - \bar{x}_s\|_2 \quad (3)$$

which is the Euclidean distance between the centroids of two cells where,

$$\bar{x}_r = \frac{1}{n_r}\sum_{i=1}^{n_r} x_{ri} \quad (4)$$

The matrix generated from this function is a *(m-1)×3* matrix, where m is the number of machines in the original dataset. Columns of the matrix contain cluster indices linked in pairs to form a binary tree. The leaf nodes are numbered from 1 to *m*. Leaf nodes are the singleton clusters from which all higher clusters are built. Further The dendrogram can be obtained from the matrix which indicates a tree of potential solutions. The hierarchical relationships and dendrogram are presented in Fig. 5 and Fig. 6.

| Node | Group 1 | Group 2 | Simil. |
|------|---------|---------|--------|
| 1    | m1      | m4      | 0.857  |
| 2    | m2      | m3      | 0.667  |
| 3    | Node 2  | m5      | 0.319  |
| 4    | Node 1  | Node 3  | -0.259 |

Fig.5 Hierarchy formation based on similarity index

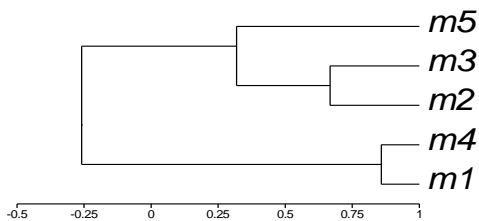

Fig.6 Dendrogram of machine grouping

## D. Part family formation technique

The following part family formation technique is adopted from reference [24] and modified substantially. Parts are assigned to the cells which further form part families using membership index given in (5). In this function the number of voids in a particular cell and number of unused machines by a part in a particular cell is also considered, which further attempts to remove a part from a part family which has larger number of unused machines in that particular cell. Therefore this function can eventually propose a trade-off between larger number of machines used for that part and also larger number of machines unused by the part in terms of performance measure criteria.

$$D_{cj} = \frac{N_{cj}}{m_c} \times \frac{N_{cj}}{n_j} \times \frac{1}{v_c} \quad (5)$$

$D_{cj}$ = Membership index of part j to cell c
$N_{cj}$ = Number of machines in cell c which process part j
$m_c$ = Total number of machines in cell c
$n_j$ = Total number of machines required by part j
$M_{cj}$ = Number of machines in cell c which do not process part j
$v_c$ = Number of voids in cell c.

Sooner the part families are formed the proposed iterative procedure investigates the grouping efficacy value in every iteration and attempts to maximize the value of grouping efficacy.

It eventually stops if the solution has not been improved after a certain number of consecutive iterations. Setting up the parameter of maximum number of iterations would be a crucial job of the decision makers since lower value could result in premature convergence and higher value could increase the computational time. In general the heuristic can execute upto maximum 100 iterations for medium to large-size problems. For an example the heuristic was executed for 40 iterations for 7×11 matrix proposed by [25], which has shown the largest number of iterations. The convergence curve is shown in Fig. 7 and the final output matrix is shown in Fig. 8.

Proposed Hybrid Algorithm

Input: *Machine-part incidence matrix A*
1. Procedure *similarity ()*
1.1. Compute similarity values between pair of machines using equation (2)
1.2. Compute the similarity matrix of the machines $S_m$
1.3. End
2. Procedure *CLCA ()*
2.1. Loop
2.2. Compute the Euclidian distance between the centroids of two cells
2.3. Construct matrix of size (m-1)×3 to from the hierarchical tree structure
2.4. Construct dendrogram from the binary matrix computed using centroid linkage rule
2.5. Loop
2.6. Create machine cells for the highest level of similarity coefficient
2.7. End
3. Procedure *part_family_formation_heuristic ()*

*3.1. Find a machine cell which processes the part for a larger number of operations than any other machine cell and assign the part in that machine cell.*
*3.2. If tie occurs, choose the machine cell which has the largest percentage of machines visited by the part and assign in that cell*
*3.3. Calculate the objective value 'f'*
*3.4. Loop*
*3.5. Select the part with larger number of unused machines in cell and assign the part to other cell*
*3.6. Check for the objective function value 'f1'*
*3.7. If the f1>f,*
*3.8. Accept the arrangement*
*3.9. Else assign the part to another cell*
*3.10. Repeat step 3.4-3.9 for every part assigned to cell*
*3.11. Stop if maximum iteration has been reached*
Output: *Optimized machine cell configuration and part family structure*

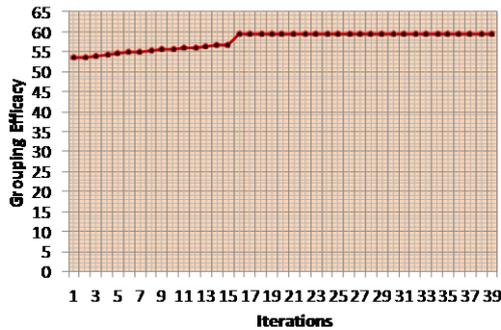

Fig. 7 Convergence curve of the heuristic for problem #3

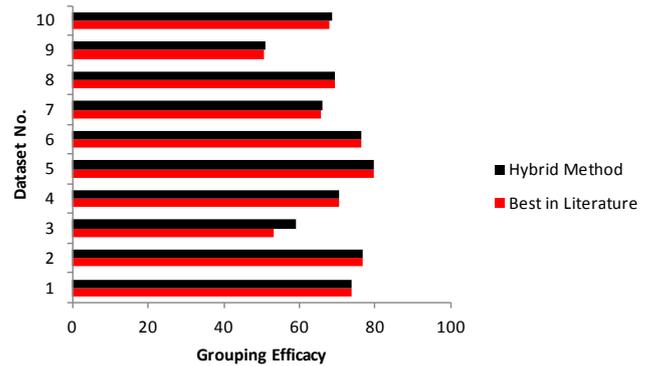

Fig.8 Output matrix

## V. COMPUTATIONAL RESULTS

The proposed algorithm is tested with a set of 10 problems that have been published in the past literature and have been widely practiced in many comparative studies. The algorithm is coded with Matlab 7.1 and executed on a laptop with a 2.1GHz processor and 2GB of RAM.

Comparisons of the proposed method against best results from the literature are given in table I. The best results are obtained from Unlar and Gungor [7]. For the problems solved with the proposed method to obtain optimal solution, the grouping efficacy value is improved or equal in all instances.

From table I it can be stated that the proposed method produces equal efficacy value in 6 instances where as in 4 instances it outperforms other established techniques. It can be quantified that the proposed technique produces 40% improved result than the best results obtained from literature. Most of the problems took negligible computational time (< 5 CPU seconds). Fig. 9 shows the performance of the proposed technique against best found methods of literature, which clearly depicts that the proposed method depicts improved performance over the other techniques.

TABLE I
COMPARISON OF THE RESULTS

| # | Reference | size | Best result in literature | Hybrid method | improvement |
|---|---|---|---|---|---|
| 1 | [23] | 5×7 | 73.68 | 73.68 | 0.00% |
| 2 | [26] | 6×8 | 76.92 | 76.92 | 0.00% |
| 3 | [25] | 7×11 | 53.13 | **59.26** | **11.54%** |
| 4 | [27] | 7×11 | 70.37 | 70.37 | 0.00% |
| 5 | [28] | 5×18 | 79.59 | 79.59 | 0.00% |
| 6 | [29] | 10×10 | 76.47 | 76.47 | 0.00% |
| 7 | [30] | 14×24 | 65.75 | **66.2** | **0.70%** |
| 8 | [31] | 14×24 | 69.33 | 69.33 | 0.00% |
| 9 | [11] | 16×24 | 50.48 | **51.04** | **1.10%** |
| 10 | [32] | 16×30 | 67.83 | **68.5** | **1%** |

Fig. 9 Comparison of proposed method with other approaches

## VI. CONCLUSIONS

This study portrays a hybrid clustering technique that combines Sorenson's similarity coefficient method with a hierarchical centroid linkage clustering technique. Computational results presented in Section 5 demonstrate that the proposed technique outperforms the best result obtained from recent literature. This article states that, the proposed method not only improves the solution quality substantially, but also reduces the variability of the solutions obtained. It attains better quality solutions by consuming lesser computational time and resources than that of the traditional methodologies. It is also shown that the proposed approach performs at least as well as, and often better than, some of the best algorithms for the cell formation on all problems tested.